\documentclass[runningheads]{llncs}
\usepackage{graphicx, subcaption}
\begin{document}
%
\setlength{\parskip}{0em}

\title{Medical Literature Mining and Retrieval in a Conversational Setting}
%
%
\newcommand*\samethanks[1][\value{footnote}]{\footnotemark[#1]} 

\author{Souvik  Das\inst{1} \thanks{equal contribution} \and
Sougata Saha \inst{1} \samethanks   \and
Rohini K. Srihari\inst{1}}
\authorrunning{S. Das et al.}
\institute{State University of New York at Buffalo, Buffalo NY, USA \\
\email{\{souvikda, sougatas, rohini\}@buffalo.edu}\\
}
\maketitle              
\begin{abstract}
The Covid-19 pandemic has caused a spur in the medical research literature. With new research advances in understanding the virus, there is a need for robust text mining tools which can process, extract and present answers from the literature in a concise and consumable way. With a DialoGPT based multi-turn conversation generation module, and BM-25 \& neural embeddings based ensemble information retrieval module, in this paper we present a conversational system, which can retrieve and answer coronavirus related queries from the rich medical literature, and present it in a conversational setting with the user. We further perform experiments to compare neural embedding based document retrieval and the traditional BM25 retrieval algorithm, and report the results.
\end{abstract}
\section{Introduction}
The ongoing Covid-19 pandemic has seen a spur in medical research. With the number of infections well over 25 million world wide, there has been a burst of research for finding a cure of the highly infectious disease. With the volume of publications related to coronaviruses increasing steeply, there is a need for robust text mining tools which can aid medical researchers to survey the existing biomedical publications effectively and exhaustively. Keeping this use case in mind, we present a question answering socialbot that can not only mine the existing medical literature, but also conduct open domain chit-chat dialogue, thus acting as a personal assistant for a researcher.
In the field of natural language processing, there has been significant amount of research in question answering(QA) systems. The introduction of transformer architecture \cite{vaswani2017attention} has propelled the QA capabilities of NLP systems. Large pre-trained transformer models like BERT \cite{Devlin_2019} have attained state of the art performance in standard open domain extractive QA datasets like SQuAD \cite{Rajpurkar_2018}, \cite{Rajpurkar_2016}. Although transformer based architectures attain state of the art performance, they have a high time complexity in scenarios where answer is to be extracted by processing multiple documents. \cite{seo-etal-2019-real} introduces a dense-sparse phrase index to alleviate the inference time bottleneck, by generating and indexing phrases from a corpus.

The Alexa prize socialbot grand challenge is testimony to the advances made in conversational systems. With the goal of creating a conversational system that can converse with a human for more than 20 minutes, and receive a user satisfaction score of 4 out of 5, there has been significant research in trying to attain this goal. The winners of the Alexa prize socialbot challenge 3 \cite{Finch2020EmoraAI},and the runners up \cite{paranjape2020neural} leverage an array of state of the art NLP techniques to develop a socialbot that converses like human. Using more sophisticated techniques for bettering the NLU(natural language understanding) module of a chatbot, the Alexa prize challenge has seen an increase in user approval ratings of chatbot conversations. \cite{Radford2019LanguageMA} introduced GPT-2, a transformer based pre-trained language model which unlike BERT can only attain to the seen context. With 1.5 billion parameters, GPT-2 achieves state of the art performance in 7 out of 8 language modelling tasks in a zero-shot setting. \cite{Zhang_2020} introduces DialoGPT, a conversational response generation model which extends GPT-2, and is trained on Reddit comment pairs.

In March 2020 Allen Institute for AI (AI2) released the first version of the COVID-19 Open Research Dataset(CORD-19) dataset \cite{Wang2020CORD19TC}, which contains all Covid-19 and coronavirus related research from PubMed's PMC, WHO(World Health Organization), bioRxiv and medRxiv pre-prints. The CORD-19 dataset has been ever increasing, and has fuelled several IR competitions in Kaggle and the TREC-Covid IR challenge \cite{10.1093/jamia/ocaa091}. In the past few months, several text mining and retrieval systems have been implemented within the NLP research community with the objective of making it easy for the medical research community to retrieve information from coronavirus related publications. \cite{oniani2020qualitative} proposed a language model for automatically answering questions related to COVID-19. They utilized the GPT-2 language model and applied transfer learning to retrain it on the CORD-19 corpus. \cite{lee2020answering} proposes covidAsk, a question answering (QA) system that combines biomedical text mining and QA techniques to provide answers to questions in real-time. They leverage both supervised and unsupervised approaches to provide informative answers using DenSPI \cite{seo-etal-2019-real} and BEST \cite{10.1371/journal.pone.0164680}. 

Inspired by the recent developments in NLP, and with the motive of enabling medical research for finding a cure for Covid-19, we present a DialoGPT based socialbot, which is not only capable of conducting open domain chat in a multi-turn setting, but also able to detect user intent and leverage a combination of traditional information retrieval models like BM25, and transformer based models like BERT and bioBERT \cite{Lee_2019}. We leverage the DailyDialog dataset \cite{li2017dailydialog} to fine tune DialoGPT on multi-turn open domain conversations. For document retrieval, we implement an ensemble of rank-bm25, a python implementation of Okapi BM25, and bioBERT embeddings. For mining and extracting the answer span from a relevant document, we leverage a pre-trained BERT large model which is fine tuned on the SQuAD dataset. In the next sections we detail our implementation and share the results. Finally we perform error analysis to understand the pitfalls of the system and propose potential future research.
\section{Methodology and Experiments}
    \subsection{Datasets}
        \subsubsection{Chatbot}
        {
            We use the dailydialogue  dataset \cite{li2017dailydialog} for fine-tuning the Dialog-GPT pretrained model. The dataset consists of manually written multi-turn dialogues, with manually labelled communication intent and emotion. The dataset consists of conversations pertaining to 10 different day to day topics like ordinary life, school life, culture \& education, attitude \& emotion, relationship,  tourism  ,  health,  work,  politics,  and finance. The dataset captures 7 types of emotion:anger, disgust, fear, happiness, sadness, surpris eand no emotion, and 4 distinct dialogue acts:  in-form, questions, directives and commissive. 
        }
        
        \subsubsection{COVID Turn Classifier}
        {
            For building the COVID text classifier we have used COQB-19 dataset \cite{COQB19}, this dataset consists of a total of 3,924 questions. Each question is assigned with a Question ID. Questions of the same ID are regarded as similar questions and grouped together in the dataset. The dataset is further divided into 2 parts, (1) Crowd-Sourced: This set consists of a total of 2,341 questions with 280 unique Question IDs. (2) Auto-generated: This set consists of a total of 1,583 questions with 68 unique Question IDs. For our experimentation we have only considered the crowd-sourced questions. The negative examples for the classifier we have randomly sampled 3593 unique turns from the daily dialog dataset.
        }
        
        \subsubsection{Information Retrieval system:}
        With the aim of providing text mining capabilities to medical researchers, we perform information retrieval and question answering on the the CORD-19 open research dataset. Specifically we use the 19\textsuperscript{th} June 2020 dump of the dataset. The CORD-19 dataset consists of the latest medical literature related to Covid-19 and coronaviruses. We further leverage the relevance judgement scores from the 4\textsuperscript{th} round of the TREC-COVID information retrieval challenge for experimenting with different retrieval systems.
        
    \subsection{Chatbot Architecture}
    
    \begin{figure}[t!]
    \centering
    \includegraphics[width=60mm]{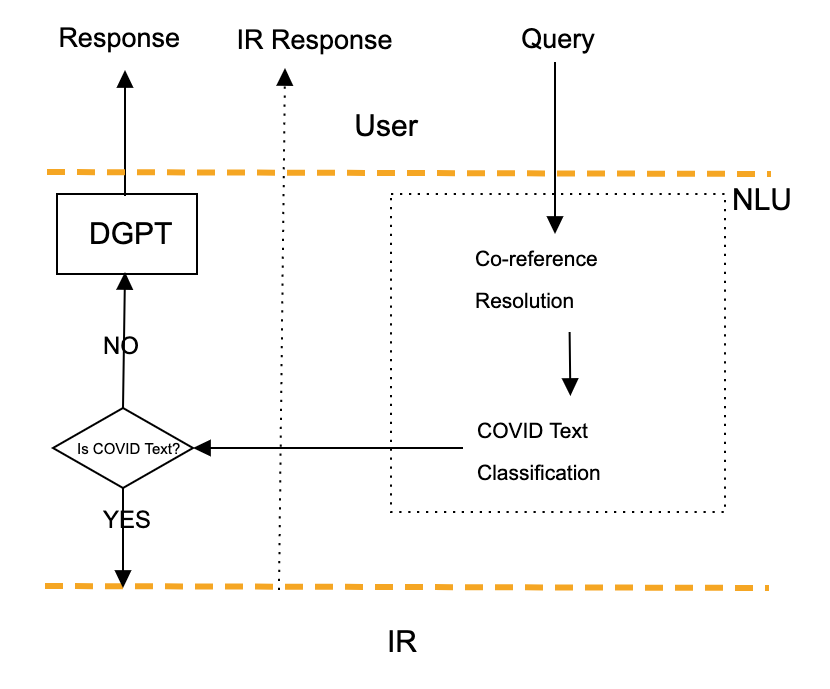}
    \caption{Architecture of chat processing and generation pipeline.} \label{fig1}
    \end{figure}
    
        \subsubsection{Dialog Generation Using DialoGPT}
        We have fine-tuned the DialoGPT pre-trained model on the Daily-Dialog dataset.  During training, we have conditioned the model to classify emotion and dialog-act of the reply.   Our model inherits the overall DialoGPT architecture, which is based on GPT2. For generating the response for general queries we employ a top-k sampling approach.
        \subsubsection{Natural Language Understanding(NLU) Module}
        NLU module ingests all the user turns and performs co-reference resolution on the most recent user turn, with the uber goal of being able to resolve COVID related terms. The resolved text is passed through a transformer based text classifier for classifying whether the text contains any Covid related text. If the text is classified as Covid text the we run this text through a keyword analyzer, which matches the text with some predefined keywords related to Covid, MERS and SARS, and enriches the query by attaching keywords related to the disease. The processed text is then passed to the IR system to fetch potential information related to the need of the user. For general queries we generate responses using the fine-tuned Dialo GPT model.

    \subsection{Information Retrieval Architecture} 
\begin{figure}[t!]
    \centering
    \includegraphics[width=0.7\textwidth]{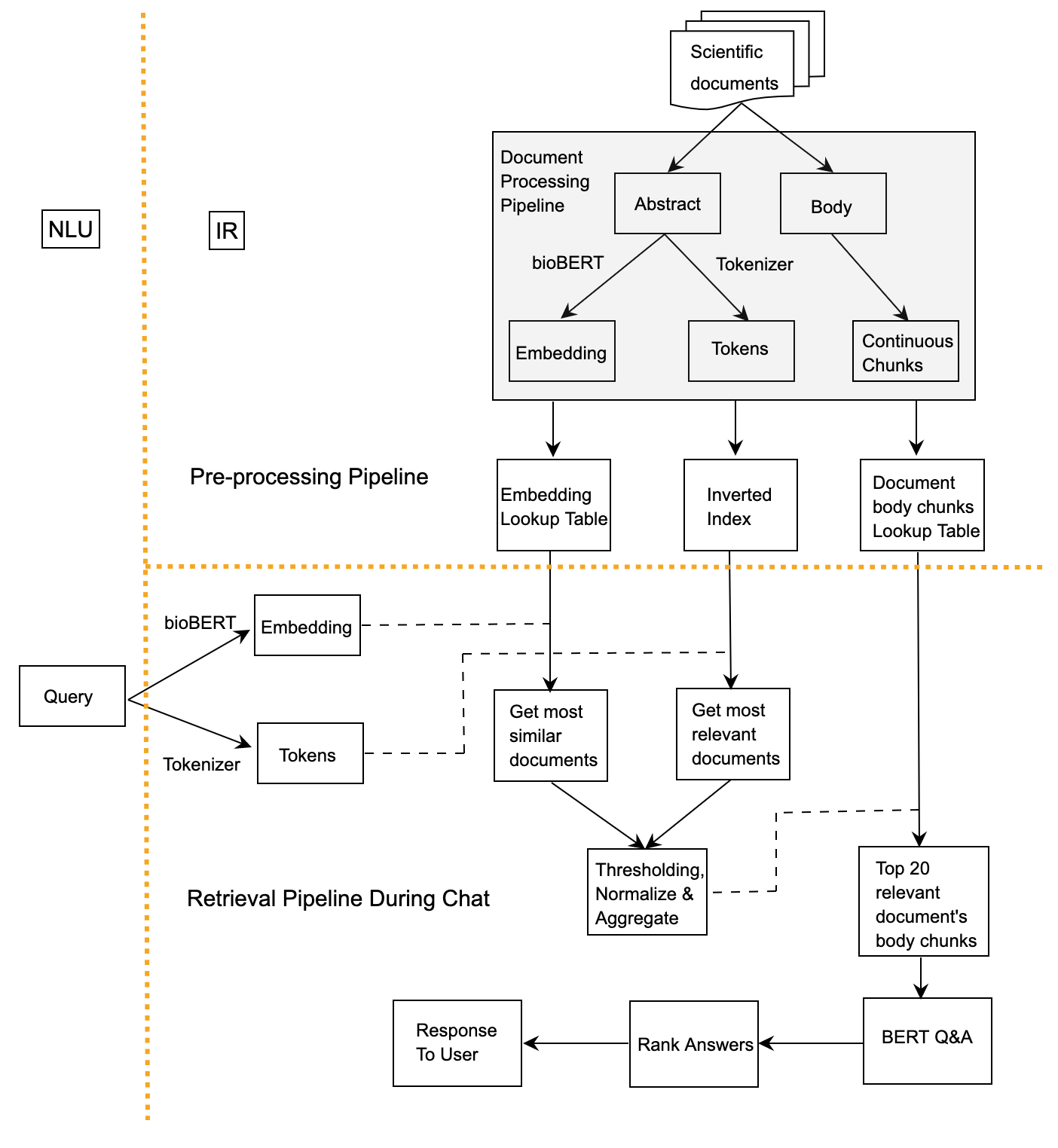}
    \caption{Architecture of the document pre-processing, and the inference time document retrieval pipeline.} \label{fig2}
    \end{figure}

We implement a hierarchical IR architecture for retrieving the correct response to a user query. We utilize an ensemble of neural embedding based and traditional IR based retrieval for retrieving top documents, which can potentially answer the user query. We perform question answering on the body text of the retrieved documents to extract the required user information, and return a ranked set of responses to the user. Fig. \ref{fig2} illustrates the document pre-processing and retrieval architecture.
    
    \subsubsection{Document pre-processing \& Indexing:} For our analysis, we only consider documents which have both an abstract and body in the CORD-19 dataset. The CORD-19 dataset contains data from multiple sources. In order to maintain the most complete representation of a document, we keep the data from the source which yields higher tokens. We further filter out documents which have either more than 300 tokens in abstract or more than 100 paragraphs of text in the body. Post processing, our document collection stands at 53,000 documents with abstract and body text present. We store the documents in 3 different ways: i) We tokenize the document abstracts and create an inverted index over the tokens. ii) We generate 768 dimensional bioBERT embeddings of the abstracts, and maintain a lookup of the document abstracts. iii) We implement a sliding window of 220 tokens with an overlap of 50 tokens, and generate continuous chunks from the document body, and maintain a lookup between the abstract inverted index and embeddings.
    \subsubsection{Document Retrieval:} Since an abstract is a succinct summary of a scientific document, we determine the relevance of a document to a query, by calculating the relevancy between the abstract and the user query. Given a user query, we simultaneously search the previously created inverted index for the most relevant abstract using the Okapi BM25 algorithm, and also calculate the cosine similarity between the bioBERT embedding of the user query and the abstract embeddings to retrieve the most similar abstracts to the query. We apply thresholds on the cosine similarity score and BM25 score to determine if a document should be considered to be relevant. The thresholds are determined by experimentation on the TREC dataset, which is discussed in section \ref{ir_exp}. Post thresholding, we normalize(between 0 and 1) and aggregate the two relevance scores, and re rank the documents by the aggregated score. We consider the top 20 ranked documents, and pass the body text to the downstream question answering module, for extracting the exact information from the relevant document.
    \subsubsection{Question \& Answering(QA): } Given a list of most relevant documents, the QA module performs BERT question answering on the chunks of body text of a document, and extracts the most relevant span. We lookup the body chunks of the top 20 documents retrieved by the document retrieval module, and use pre-trained BERT large model which is fine tuned on the SQUAD dataset to perform question answering on each of the body chunks. Given a query and a text passage, the BERT question answering model generates the log likelihood of each of the tokens in the passage, as being a start and end token in the answer span. We extract the most probable span from a text chunk, and leverage the log likelihood of the starting token of the span for ranking and filtering the answers, which is discussed below.
    \subsubsection{Ranking \& Response Generation: } We apply the following filtration criteria to the predicted spans to check their eligibility of being an answer. 
    i) We observed that longer spans were associated with non favourable answers, hence applied a criteria that the span should not be more than 15 tokens. ii) The answer should not contain any special BERT specific tokens like `CLS', `SEP' and `PAD'. Since we perform question answering on multiple chunks of the same document, we select one answer per relevant document by selecting the answer whose start token has the highest log likelihood. Finally, we rank the answers from the relevant documents based on the start token log likelihood and the normalized \& aggregated BM25 and cosine similarity score, and yield the top 5 answers along with the document name(paper name) to the user. If none of the answers pass the filtration criteria, we return a ranked list of 5 most relevant document names(paper names) to the user based on the document retrieval score(BM25 + cosine similarity score). In the case of no document being retrieved by the retrieval module, we respond with a response seeking clarification from the user.
    \section{Experiments}
    \subsection{Fine Tuning Dialo-GPT}{
        The model has 125M trainable parameters, and we have trained the model using a training batch size of 32, gradient accumulation step of 8 and a learning rate of 1e-5. We have used Adam optimizer \cite{kingma2014adam} with max grad norm set to 1. The model was trained for 5000 optimization steps with 2000 warmup steps. The weight-decay for all model parameters except for biases and LayerNorm \cite{ba2016layer} was set to 0.01.
    }
    \subsection{COVID Turn Classifier}
    {
        We trained a transformer based classification model using the COQB-19 dataset for the positive examples and randomly sampling DailyDialog utterances as negative examples. The model consists of a transformer layer followed by a linear layer to make a binary prediction of whether the text is coronavirus related or open domain. For our experimentation we have used BERT base pre-trained model, we apply dropout with a probability of 0.3 before passing through the linear layer. A softmax layer was used to calculate the class probabilities. Loss is calculated using binary cross entropy function. The entire dataset was split into 70:15:15 ratio for training, validation and test sets respectively. The model is trained for 2 epochs with maximum length of the text equal to 250 characters, batch size of 32 and learning rate of 2e-5. We have used Adam optimizer with max grad norm set to 1.
    }
    \subsection{IR System}
    \label{ir_exp}
    We compared 3 different retrieval strategies: i) Using Okapi BM-25 for scoring and retrieving relevant document abstracts. ii) Cosine similarity between bioBERT query and abstract embedding. iii) Using Okapi BM-25 score and embedding cosine similarity and retrieving the union of the two strategies. Between April to July 2020, TREC conducted an information retrieval challenge, where the task was to retrieve the most relevant documents from the CORD-19 dataset, for a set of 45 Covid-19 related queries. They also provided a set of human judged query relevance judgement scores(TREC paper relevance judgement). The relevance judgement scores indicated whether a document was highly relevant, mildly relevant or not relevant to a query. In order to determine the optimal document retrieval strategy, we leveraged the relevance judgement scores and performed a comparison study. We converted the three class problem to binary problem (relevant or not relevant) by collapsing the highly and mildly relevant classes into one, and tested our document retrieval strategies by comparing the F1 score. Since BM25 and cosine similarity generates scores and not classes, we had to determine thresholds on the scores to classify a document as relevant or not relevant, and then calculate the F1 score. For each retrieval strategy, we performed an extensive grid search to determine the optimal threshold.
    \section{Results}
    We attain a perplexity score of 13.125 by jointly fine tuning DialoGPT to generate responses as well as predicting the dialogue act and emotion of the turns. Mathematically perplexity it is defined as the exponentiation of entropy ($2^{H(p)}$), and a lower score is preferred. Table \ref{tabcl} shows the test set performance of the Covid Turn Classifier. Though we observe a high performance, the model fails to understand a few ambiguous cases when a speaker talks about a random virus. We believe implementing a way to better negative samples will yield better results.
    
    We report the results of the different document retrieval strategies in Table \ref{tab1}. We observe that retrieving documents based on the cosine similarity of the abstract bioBERT embedding performs worse than BM25 based retrieval. We reason that although embedding based approaches are superior for many NLP tasks, where it is important to represent near similar or semantically similar documents nearby in the embedding space, in search the presence of exact keywords matters more. We further notice that retrieving documents based on a union of embedding cosine similarity and BM25 score yields best results, as this captures the best of both retrieval methods.
        \begin{table}[t!]
            \centering
            \caption{Testing performance of the COVID Turn Classifier}\label{tabcl}
            \begin{tabular}{|l|l|}
            \hline
            Evaluation Metric   & Score    \\ \hline
            Accuracy & 0.9833 \\ 
            F1       & 0.9850 \\ \hline
            \end{tabular}
            \end{table}
    
    \begin{table}[t!]
    \centering
    \caption{Comparison of F1 score between different IR approaches.}\label{tab1}
    \begin{tabular}{|l|l|l|}
    \hline
    Retrieval Method &  F1 & Threshold\\
    \hline
    Okapi BM 25 &  0.6527 & 2.77\\
    Cosine Similarity &  0.6315 & 0.89\\
    Okapi BM 25 + Cosine Similarity & {\bfseries 0.6713} & 2.77, 0.89\\
    \hline
    \end{tabular}
    \end{table}

\section{Discussion}
    
    \begin{figure}[t]
      \centering
      \begin{subfigure}{.5\linewidth}
        \centering
        \includegraphics[width=\textwidth]{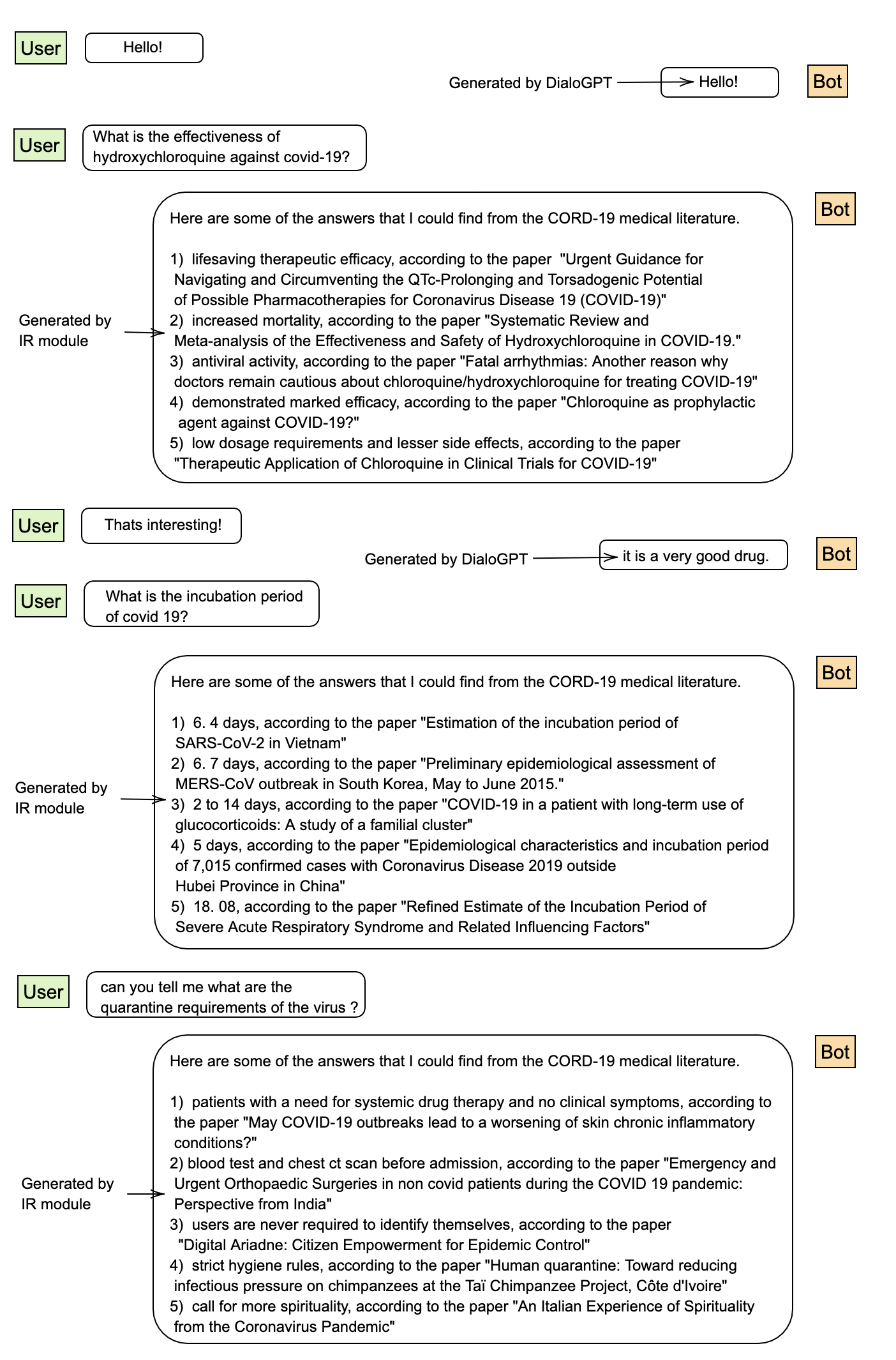}
        \caption{Actual chat conducted with the conversational system.}
      \end{subfigure}%
      \hfill
      \begin{subfigure}{.5\linewidth}
        \centering
        \includegraphics[width=\textwidth]{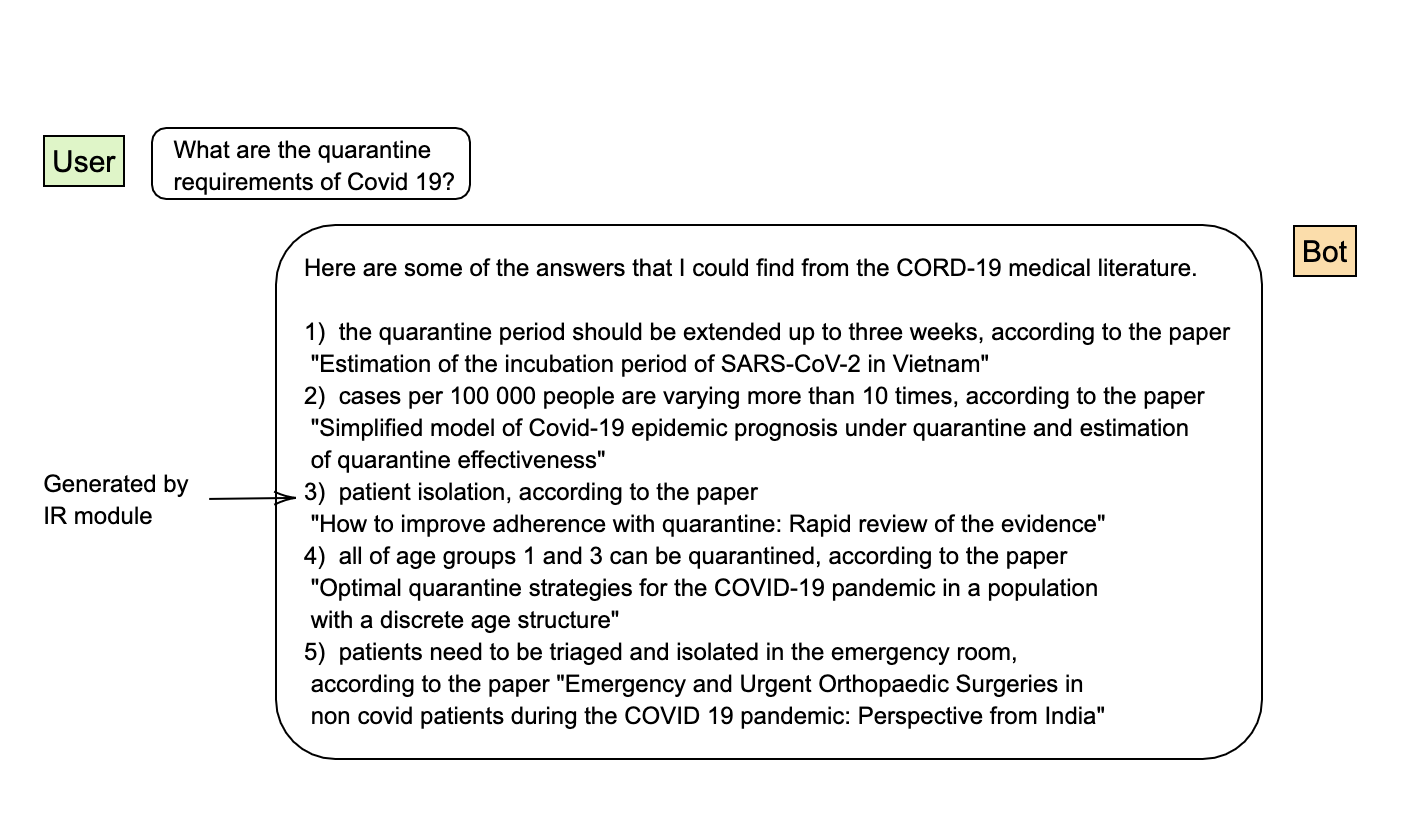}
        \caption{Correct documents retrieved by the IR module with a reformulated query. Note how the last user query in chat (a) retrieves non-relevant documents.}
      \end{subfigure}%
      \caption{Sample of chats conducted with the conversational system. Please zoom in for better viewability.}
      \label{chat_example}
    \end{figure}
    
    Fig.\ref{chat_example} depicts an example chat conducted with the conversational system. We have annotated the chatbot responses with the module that was responsible for the reply. Keeping the DialoGPT based response generation module aside, there are 3 main kinds of errors related to information retrieval that the system makes: i) Retrieval error: When the IR system retrieves a wrong document. As shown in the experiments, a ensemble approach of embedding based and traditional based retrieval works better, which can be made more efficient by implementing a hierarchical index of the document body text instead of the abstract. ii) Answer span selection error: When the BERT QA model predicts spurious span as answer to a query. This error can be reduced by creating a question answering dataset for the medical literature, and fine tuning BERT for question answering in a supervised learning setting. iii) Query formulation error: When the query is ill formed by the user or the NLU, which can result in retrieval of the wrong documents. This can be reduced by reformulating the user query to adhere to a specific template which can be passed to the IR system. Fig. \ref{chat_example} depicts a similar scenario.
    
\section{Conclusion}   
    In this paper we present a document retrieval and text mining system, which can mine the biomedical scientific literature for answers, and have implemented a GPT-2 based conversational system as an interface for interacting with a user. We use a natural language understanding unit for deciding whether to send a query to the IR system for retrieving answers from the biomedical literature, or to generate an open domain response based on the previous context of the conversation. In future, we intend to implement a paraphrasing and natural text generation module, which can take as input the most relevant chunk of body text, and generate a response in a more conversational way.



%
%
%
\bibliographystyle{splncs04}
\bibliography{references}

\begin{thebibliography}{10}
\providecommand{\url}[1]{\texttt{#1}}
\providecommand{\urlprefix}{URL }
\providecommand{\doi}[1]{https://doi.org/#1}

\bibitem{COQB19}
Coqb-19: Covid-19 question bank,
  \url{https://www.newvoicesnasem.org/data-downloads}

\bibitem{ba2016layer}
Ba, J.L., Kiros, J.R., Hinton, G.E.: Layer normalization (2016)

\bibitem{Devlin_2019}
Devlin, J., Chang, M.W., Lee, K., Toutanova, K.: Proceedings of the 2019
  Conference of the North  (2019). \doi{10.18653/v1/n19-1423},
  \url{http://dx.doi.org/10.18653/v1/N19-1423}

\bibitem{Finch2020EmoraAI}
Finch, J.D., Ahmadvand, A., Dong, X., Qi, R., Sahijwani, H., Volokhin, S.,
  Wang, Z., Wang, Z., Choi, J.D.: Emora: An inquisitive social chatbot who
  cares for you (2020)

\bibitem{kingma2014adam}
Kingma, D.P., Ba, J.: Adam: A method for stochastic optimization (2014)

\bibitem{lee2020answering}
Lee, J., Yi, S.S., Jeong, M., Sung, M., Yoon, W., Choi, Y., Ko, M., Kang, J.:
  Answering questions on covid-19 in real-time (2020)

\bibitem{Lee_2019}
Lee, J., Yoon, W., Kim, S., Kim, D., Kim, S., So, C.H., Kang, J.: Biobert: a
  pre-trained biomedical language representation model for biomedical text
  mining. Bioinformatics  (Sep 2019). \doi{10.1093/bioinformatics/btz682},
  \url{http://dx.doi.org/10.1093/bioinformatics/btz682}

\bibitem{10.1371/journal.pone.0164680}
Lee, S., Kim, D., Lee, K., Choi, J., Kim, S., Jeon, M., Lim, S., Choi, D., Kim,
  S., Tan, A.C., Kang, J.: Best: Next-generation biomedical entity search tool
  for knowledge discovery from biomedical literature. PLOS ONE
  \textbf{11}(10),  1--16 (10 2016). \doi{10.1371/journal.pone.0164680},
  \url{https://doi.org/10.1371/journal.pone.0164680}

\bibitem{li2017dailydialog}
Li, Y., Su, H., Shen, X., Li, W., Cao, Z., Niu, S.: Dailydialog: A manually
  labelled multi-turn dialogue dataset (2017)

\bibitem{oniani2020qualitative}
Oniani, D., Wang, Y.: A qualitative evaluation of language models on automatic
  question-answering for covid-19 (2020)

\bibitem{paranjape2020neural}
Paranjape, A., See, A., Kenealy, K., Li, H., Hardy, A., Qi, P., Sadagopan,
  K.R., Phu, N.M., Soylu, D., Manning, C.D.: Neural generation meets real
  people: Towards emotionally engaging mixed-initiative conversations (2020)

\bibitem{Radford2019LanguageMA}
Radford, A., Wu, J., Child, R., Luan, D., Amodei, D., Sutskever, I.: Language
  models are unsupervised multitask learners (2019)

\bibitem{Rajpurkar_2018}
Rajpurkar, P., Jia, R., Liang, P.: Know what you don’t know: Unanswerable
  questions for squad. Proceedings of the 56th Annual Meeting of the
  Association for Computational Linguistics (Volume 2: Short Papers)  (2018).
  \doi{10.18653/v1/p18-2124}, \url{http://dx.doi.org/10.18653/v1/P18-2124}

\bibitem{Rajpurkar_2016}
Rajpurkar, P., Zhang, J., Lopyrev, K., Liang, P.: Squad: 100,000+ questions for
  machine comprehension of text. Proceedings of the 2016 Conference on
  Empirical Methods in Natural Language Processing  (2016).
  \doi{10.18653/v1/d16-1264}, \url{http://dx.doi.org/10.18653/v1/D16-1264}

\bibitem{10.1093/jamia/ocaa091}
Roberts, K., Alam, T., Bedrick, S., Demner-Fushman, D., Lo, K., Soboroff, I.,
  Voorhees, E., Wang, L.L., Hersh, W.R.: {TREC-COVID: rationale and structure
  of an information retrieval shared task for COVID-19}. Journal of the
  American Medical Informatics Association  (07 2020).
  \doi{10.1093/jamia/ocaa091}, \url{https://doi.org/10.1093/jamia/ocaa091},
  ocaa091

\bibitem{seo-etal-2019-real}
Seo, M., Lee, J., Kwiatkowski, T., Parikh, A., Farhadi, A., Hajishirzi, H.:
  Real-time open-domain question answering with dense-sparse phrase index. In:
  Proceedings of the 57th Annual Meeting of the Association for Computational
  Linguistics. pp. 4430--4441. Association for Computational Linguistics,
  Florence, Italy (Jul 2019). \doi{10.18653/v1/P19-1436},
  \url{https://www.aclweb.org/anthology/P19-1436}

\bibitem{vaswani2017attention}
Vaswani, A., Shazeer, N., Parmar, N., Uszkoreit, J., Jones, L., Gomez, A.N.,
  Kaiser, L., Polosukhin, I.: Attention is all you need (2017)

\bibitem{Wang2020CORD19TC}
Wang, L.L., Lo, K., Chandrasekhar, Y., Reas, R., Yang, J., Eide, D., Funk, K.,
  Kinney, R.M., Liu, Z., Merrill, W., Mooney, P., Murdick, D., Rishi, D.,
  Sheehan, J., Shen, Z., Stilson, B., Wade, A., Wang, K., Wilhelm, C., Xie, B.,
  Raymond, D., Weld, D.S., Etzioni, O., Kohlmeier, S.: Cord-19: The covid-19
  open research dataset. ArXiv  (2020)

\bibitem{Zhang_2020}
Zhang, Y., Sun, S., Galley, M., Chen, Y.C., Brockett, C., Gao, X., Gao, J.,
  Liu, J., Dolan, B.: Dialogpt: Large-scale generative pre-training for
  conversational response generation. Proceedings of the 58th Annual Meeting of
  the Association for Computational Linguistics: System Demonstrations  (2020).
  \doi{10.18653/v1/2020.acl-demos.30},
  \url{http://dx.doi.org/10.18653/v1/2020.acl-demos.30}

\end{thebibliography}
%





\end{document}